\documentclass{emulateapj}
\usepackage{graphicx}

\slugcomment{Submitted to ApJ}
\shorttitle{Ly$\alpha$ emission around $z>6$ quasars}
\shortauthors{Decarli et al.}

\def\Lsun{L$_\odot$}
\def\Msun{M$_\odot$}

\def\Mgii{Mg\,{\sc ii}}

\def\Cii{[C\,{\sc ii}]}

\def\Lya{Ly$\alpha$}
\def\Ha{H$\alpha$}

\def\kms{km\ s$^{-1}$}
\def\lsim{\mathrel{\rlap{\lower 3pt \hbox{$\sim$}} \raise 2.0pt \hbox{$<$}}}
\def\gsim{\mathrel{\rlap{\lower 3pt \hbox{$\sim$}} \raise 2.0pt \hbox{$>$}}}

\begin{document}

\title{HST narrow-band search for extended Ly$\alpha$ emission around two $z>6$ quasars}

\author{
Roberto Decarli\altaffilmark{1},
Fabian Walter\altaffilmark{1},
Yujin Yang\altaffilmark{1},
Chris L. Carilli\altaffilmark{2,3},
Xiahoui Fan\altaffilmark{4},
Joseph F. Hennawi\altaffilmark{1},
Jaron Kurk\altaffilmark{5},
Dominik Riechers\altaffilmark{6},
Hans-Walter Rix\altaffilmark{1},
Michael A. Strauss\altaffilmark{7},
Bram P. Venemans\altaffilmark{1}
}
\altaffiltext{1}{Max-Planck Institut f\"{u}r Astronomie, K\"{o}nigstuhl 17, D-69117, Heidelberg, Germany}
\altaffiltext{2}{NRAO, Pete V.\,Domenici Array Science Center, P.O.\, Box O, Socorro, NM, 87801, USA}
\altaffiltext{3}{Astrophysics Group, Cavendish Laboratory, JJ Thomson Avenue, Cambridge CB3 0HE, United Kingdom}
\altaffiltext{4}{Steward Observatory, University of Arizona, 933 North Cherry Avenue, Tucson, AZ 85721, USA}
\altaffiltext{5}{Max-Planck-Institut f\"{u}r Extraterrestrische Physik, Gie\ss{}enbachstra\ss{}e 1, D-85748 Garching, Germany}
\altaffiltext{6}{Astronomy Department, Caltech, 1200 East California boulevard, Pasadena, CA 91125, USA}
\altaffiltext{7}{Department of Astrophysical Sciences, Princeton University, Peyton Hall, Ivy Lane, Princeton, NJ 08544, USA}

\email{decarli@mpia.de}

\begin{abstract}
We search for extended \Lya{} emission around two $z>6$ quasars, SDSS 
J1030+0524 ($z=6.309$) and SDSS J1148+5251 ($z=6.419$) using WFC3 
narrow-band filters on board the {\em Hubble Space Telescope}. 
For each quasar, we collected two deep, narrow-band images, one sampling 
the \Lya{}  line+continuum at the quasar redshifts and one of the 
continuum emission redwards of the line. After carefully modeling the 
Point Spread Function, we find no evidence for extended \Lya{} emission. 
These observations set 2-$\sigma$ limits of $L$(\Lya, extended) 
$<3.2\times10^{44}$ erg\,s$^{-1}$ for J1030+0524 and $L$(\Lya, extended) 
$<2.5\times10^{44}$ erg\,s$^{-1}$ for J1148+5251. Given the star formation 
rates typically inferred from (rest-frame) far--infrared measurements
of $z\sim6$ quasars, these limits are well below the intrinsic bright \Lya{} emission
expected from the recombination of gas photoionized by the quasars or by the
star formation in the host galaxies, and point towards significant \Lya{} 
suppression or dust attenuation. However, small extinction values have been 
observed along the 
line of sight to the nuclei, thus reddening has to be coupled with other 
mechanisms for \Lya{} suppression (e.g., resonance scattering). No \Lya{} 
emitting companions are found, down to a 5-$\sigma$ 
sensitivity of $\sim 1\times10^{-17}$ erg s$^{-1}$ cm$^{-2}$ arcsec$^{-2}$
(surface brightness) and $\sim5\times10^{-17}$ erg s$^{-1}$ cm$^{-2}$ (assuming
point sources). 
\end{abstract}

\section{Introduction}

The host galaxies of very high-$z$ quasars ($z>6$), harboring $>10^9$ 
\Msun{} black holes, are thought to reside in the highest density 
peaks in the universe \citep[e.g.,][]{volonteri06}. Abundant cold gas 
reservoirs are necessary to feed the black hole growth in such a short time 
(the universe at $z=6$ is less than $1$ Gyr old). Such gas reservoirs would
also likely be sites of extensive star formation. Studying host galaxies 
of quasars at $z\sim6$ is therefore one way to study the
build-up of the first massive galaxies. 

Indeed, a large fraction (30--50 \%) of the $z>5$ quasars observed at 
(sub-)mm wavelengths have been detected, revealing far-infrared (FIR)
luminosities  $5\times10^{12}-2\times10^{13}$ \Lsun{} 
\citep{priddey03,bertoldi03,wang08,wang11,leipski10}. The spectral energy 
distributions of these objects suggest that star formation (rather than
black hole accretion) is powering dust heating 
\citep{beelen06,leipski10,wang11}. The associated star 
formation rates (SFRs) easily exceed several hundred \Msun{}\,yr$^{-1}$. 
Such high star formation rates are in agreement with the detection of 
bright \Cii{}$_{158\,\mu{\rm m}}$ emission that is extended on kpc scales 
\citep[e.g.][]{maiolino05,walter09}. 
Similarly, direct evidence for significant molecular gas reservoirs, 
exceeding $10^{10}$ \Msun{}, in $z\sim6$ quasar host galaxies has now been 
firmly established through observations of the redshifted CO emission
\citep{bertoldi03,walter03,walter04,carilli07,wang07,wang11,riechers09}. 

Despite these increasing observational constraints, several questions remain
unanswered: How do quasar host galaxies accrete their gas? Is the gas 
dynamically cold, and accreting through filaments 
\citep[see, e.g.,][]{haiman01,dekel09,dubois12,dimatteo12}? 
Are host galaxies severely 
obscured? What is the escape fraction of UV and \Lya{} photons produced in 
these supposedly huge star formation events \citep[][]{dayal09}?

Key information to address these questions may come from the detection
of extended UV and \Lya{} emission around high-$z$ quasars, in particular 
the luminosity, physical extent and morphology of their host galaxies and 
halos. Extended \Lya{} emission around radio galaxies and low-$z$ quasars 
has been reported in the literature 
\citep[e.g.,][]{reuland03,weidinger05,francis06,christensen06,barrio08,smith09}; 
however, so far deep observations have only been performed for one $z\sim6$ 
quasar, J2329-0301 \citep{goto09,goto12,willott11}, using ground-based
imaging and spectroscopic observations. 

The present study aims to detect extended \Lya{} emission around two
$z>6$ quasars (for which suitable narrow-band filters exist), SDSS 
J103027.10+052455.0 \citep[][$z=6.309$; hereafter, J1030+0524]{fan01} and 
SDSS J114816.64+525150.3 \citep[][$z=6.419$; hereafter, J1148+5251]{fan03}. 
The unique angular resolution offered by the {\em Hubble Space Telescope} 
(HST) allows us to disentangle the unresolved quasar light from any extended
emission. We use the new narrow-band imaging capabilities offered by the 
Wide Field Camera 3 (WFC3) in order to sample both the pure continuum (`OFF' 
images) redwards of \Lya{} and the \Lya{} + continuum (`ON' images). Through 
accurate modeling of the PSF and its uncertainties, we will be able to 
constrain the presence of extended emission around the target quasars.

Throughout the paper we will assume a standard cosmology model with 
$H_0=70$ \kms{} Mpc$^{-1}$, $\Omega_{\rm m}=0.3$ and 
$\Omega_{\Lambda}=0.7$.

\section{WFC3 observations}
\begin{figure}
\includegraphics[width=\columnwidth]{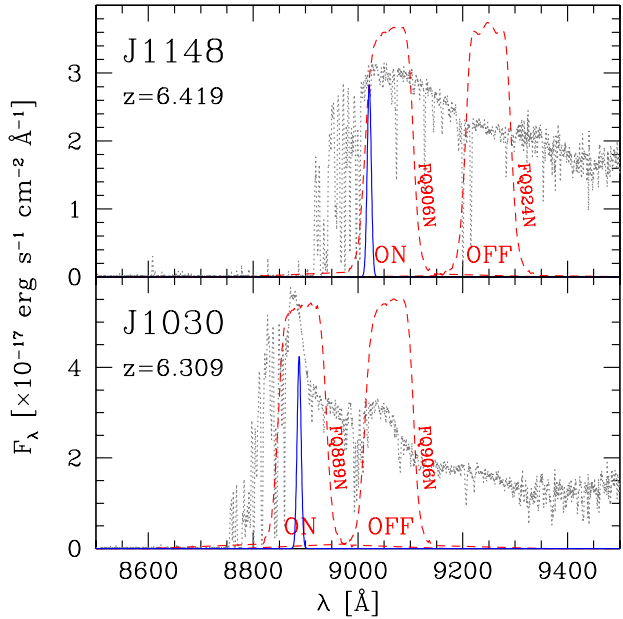}\\
\caption{Spectra of J1030+0524 and J1148+5251 (grey, dotted lines) compared 
with the throughput curves of the filters used in our analysis (red, dashed 
lines) and with the expected profile of a 300 \kms{}-broad \Lya{} line 
arising from the host galaxy (blue, solid lines). The redshift of the host 
galaxy is accurately defined by CO and \Cii{} observations for J1148+5251, 
and by the quasar \Mgii{} line for J1030+0524. Spectra of the two quasars 
are taken from \citet{pentericci02} and \citet{fan03}.}\label{fig_filters}
\end{figure}

We use the `quad-filters', i.e., a set of four different narrow-band
filters, each covering simultaneously about 1/6 of the WFC3/UVIS field of 
view ($\sim1$ arcmin$^2$). For J1030+0524 (J1148+5251), we 
used the FQ889N (FQ906N) filter for the ON images and FQ906N (FQ924N) for 
the OFF images. Figure \ref{fig_filters} illustrates the throughput 
curves of the adopted filters, the spectra of the quasars and the redshift of
the predicted \Lya{} emission from their host galaxies. The redshift is 
accurately defined by the CO and \Cii{} redshift of the source in the case 
of J1148+5251 and by the MgII line for J1030+0524.

Observations were carried out during HST Cycle 17 (proposal ID: 11640).
J1030+0524 was observed in two complete Observing Blocks (OBs; executed on
2010-01-28 and 2011-01-15) in both the ON and OFF setups (total integration 
time per OB: 5660 s in the ON setup, 5633 s in the OFF setup). During
each OB, the ON and OFF observations were performed subsequently, in a 
ON-OFF-OFF-ON sequence. J1148+5251 was observed once (2011-03-06) in the 
ON and OFF setups (total integration time: 6183 s for the ON setup, 
6167 s for the OFF setup).

Our analysis is based on data products delivered by the HST pipeline. 
Photometry is defined following the WFC3 
handbook\footnote{\textsf{http://www.stsci.edu/hst/wfc3/phot\_zp\_lbn}}. 
Theoretical zero points in the AB system are computed based on the 
\textsf{PHOTFLAM} and \textsf{PHOTPLAM} keywords and further corrected to 
account for the deviations from on-sky to theoretical zero points. 

\begin{figure}
\includegraphics[width=\columnwidth]{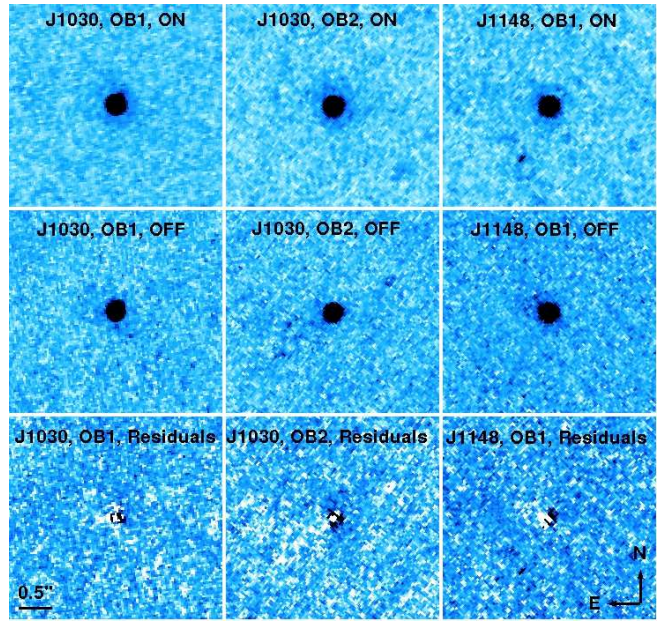}
\caption{Line+continuum (ON, top row), continuum (OFF, middle row) and 
residual images of J1030+0524 and J1148+5251, in the three OBs used in our 
analysis. Residual images are obtained as the difference between ON and OFF 
frames, after scaling the latter in order to match the flux of the former in 
the central 5$\times$5 pixels. The stretch of the color scale is linear. The 
scale and orientation of the images are the same in all the panels, as
labeled in lower left, lower right, respectively. For a
comparison, the starburst traced by \Cii{} emission observed by 
\citet{walter09} in J1148+5251 has a physical extension of $\sim0.3''$, 
while the molecular gas is distributed on scales of $\sim0.5''$ 
\citep{walter04}. No significant pure-line emission is detected around
the quasar PSFs.}\label{fig_ima}
\end{figure}

Figure \ref{fig_ima} shows the pipeline-reduced images of J1030+0524 and 
J1148+5251 in the three OBs used in this study. We do not stack the two OBs 
available for J1030+0524, in order to preserve the PSF
properties in the two observations (collected on different dates), and
to have a better control of the noise properties of the background.

The measured aperture magnitudes of the two quasars in all the OBs are 
consistent within 0.2 mag with the expected fluxes as derived from the 
spectroscopy (see Figure \ref{fig_filters}). This small difference is likely
due to absolute flux calibration uncertainties, slit flux losses in the 
spectra and intrinsic quasar variability.

\section{PSF models}\label{sec_psf}
\begin{figure*}
\includegraphics[width=\columnwidth]{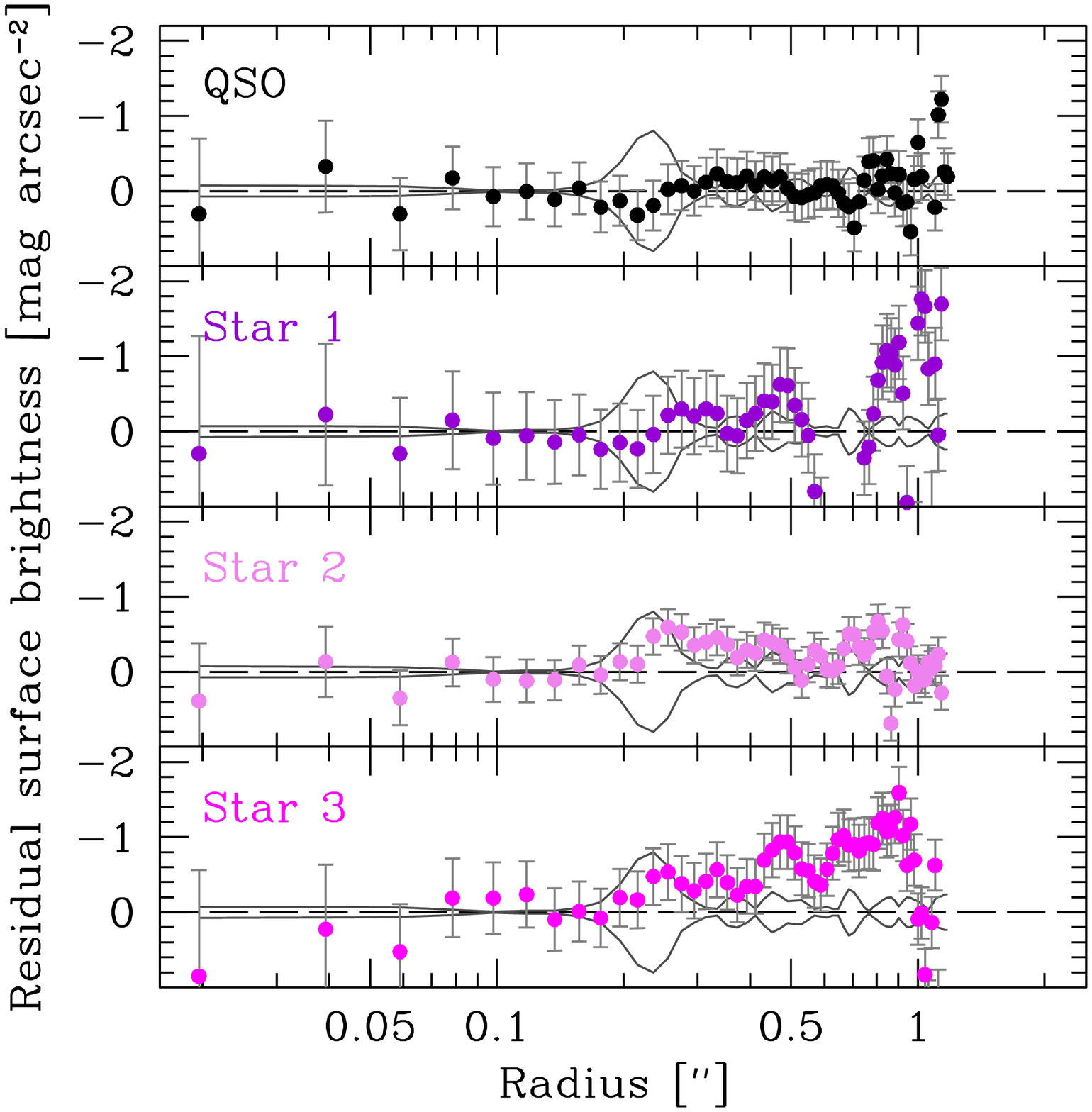}
\includegraphics[width=\columnwidth]{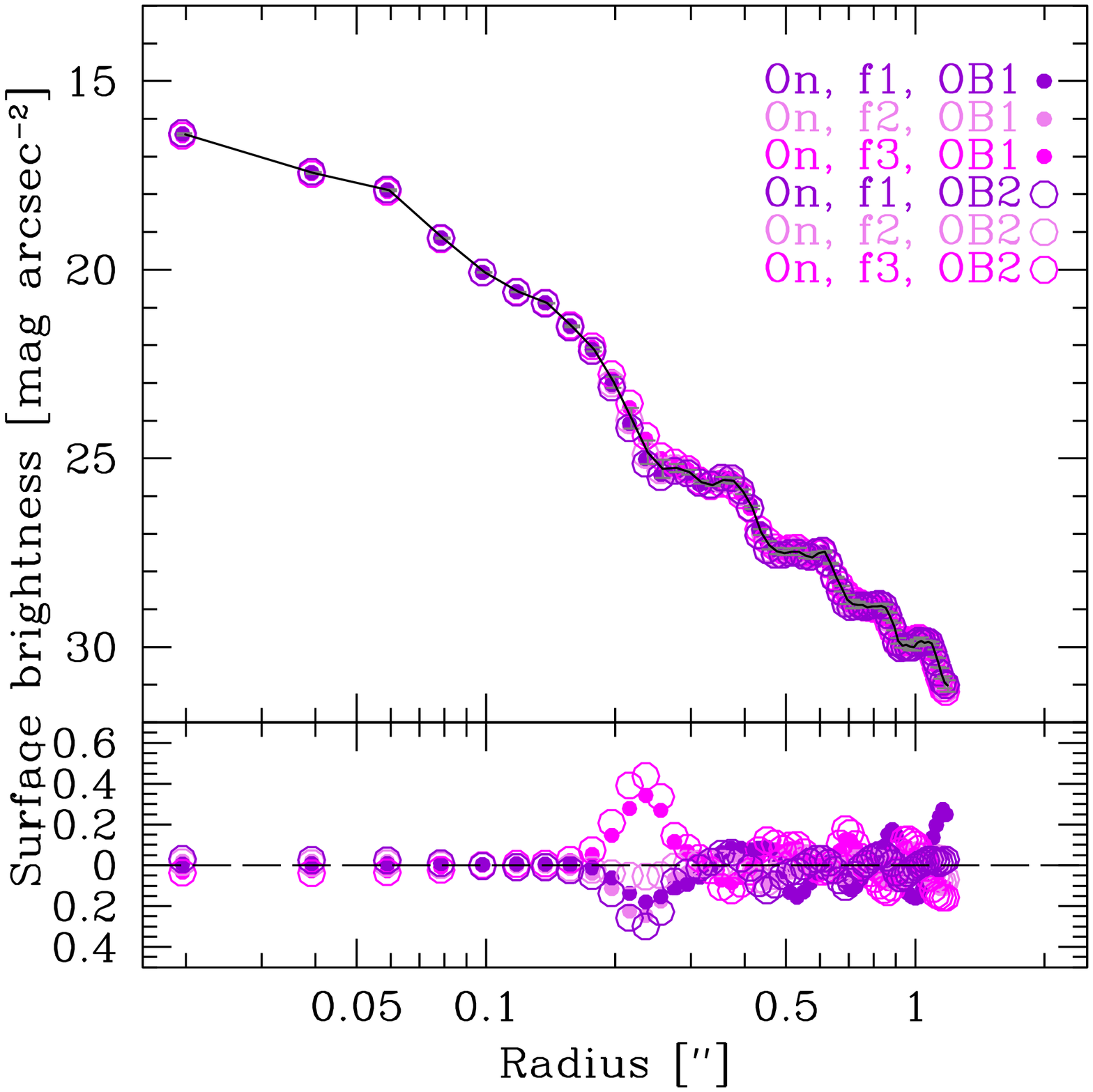}\\
\caption{PSF variations due to field degradation {\em left panel}
and HST breathing (``focus''; {\em right panel}) for J1030+0524, in the 
FQ889N filter. {\em Left:} Difference between the PSF-subtracted light 
profiles of J1030+0524 and other unresolved sources in its field. 
Significant variations are observed. The PSF model is chosen to reproduce a 
point source at the quasar position, while the field stars are $16''$, 
$44''$ and $52''$ from it, respectively. Solid lines show PSF model 
uncertainties (see section \ref{sec_psf} for details). {\em Right:} 
Radial profiles of the Tiny Tim PSF models corresponding to the average 
(f1), the maximum (f2) and the minimum  (f3) focus values. In the bottom 
panel, the residuals after the 
subtraction of the average PSF are shown.}\label{fig_psf}
\end{figure*}

In order to put constraints on extended \Lya{} emission in our targets, 
we need to model the dominant emission due to the central 
(unresolved) quasar. A common practice in quasar host galaxy studies is to 
model the quasar emission based on the images of foreground stars in the 
field \citep[see, e.g.,][]{kotilainen09}. This approach is sensitive
to spatial variations of the PSF across the field.
In order to evaluate these variations, we
compare the radial profile of three stars in the field of J1030+0524 (OB2, 
ON; the field of J1148+5251 does not contain suitable stars and therefore 
cannot be used for this experiment). Figure \ref{fig_psf}, {\em left} shows
the surface brightness profiles of these sources after the subtraction of a 
common Tiny Tim PSF model centered at the quasar position. The PSF quality 
degrades (i.e., PSF wings are more prominent) as the distance from the 
quasar position increases. Thus the PSF of field stars cannot be used as a 
model for the quasar PSF. 

Alternatively, one can use the OFF frames (including only the continuum
emission) to model the quasar image in the ON images. The major advantage
here is that the frames are always centered on the target (i.e., field
variations of the PSF are minimized) and observations are carried out
with the same focus conditions. However, this approach relies on the 
hypothesis that the quasar host galaxy does not show any extended emission
in the continuum, which is an assumption we first need to test.

We therefore use HST PSF models as simulated using Tiny Tim.
According to the WFC3 handbook, 
the PSF Full Width at Half Maximum (FWHM) shows $\sim$0.3\% variations as a 
function of the HST ``breathing'' (i.e., focus variations due to various 
causes, including thermal expansion of the satellite) at $800$ nm, 
decreasing with increasing wavelength. 
According to focus variation models\footnote{\textsf{http://www.stsci.edu/hst/observatory/focus/FocusModel/\#5}},
the focus changed significantly during the execution of the OBs.
Nevertheless, variations of the PSF at these wavelengths are limited: Figure
\ref{fig_psf}, {\em right} compares the radial light 
profile of the model PSF at the average, highest and lowest values of the 
focus during the execution of the various OBs. The most important variations
appear at aperture radii between $0.2''$ and $0.3''$. 
From these models, we adopt the PSF models that best match the observed quasar 
profile at these radii. 

PSF uncertainties shown in Fig.~\ref{fig_psf} are defined as the quadrature
sum of the PSF variations due to focus fluctuations and formal uncertainties in 
the PSF profile due to pixelization, Poissonian errors and background rms.
%

\section{Results}

In this section we describe how we use these data to search for \Lya{} 
emission arising from
the host galaxies (Sec.~\ref{sec_host}), from any filamentary structure 
around the quasars (Sec.~\ref{sec_accretion}) and from possible companion 
sources (Sec.~\ref{sec_companions}). We then present the results for our two
sources.

\subsection{Extended \Lya{} emission in the host galaxies}\label{sec_host}

In order to investigate the presence of any extended emission arising from
the host galaxies of our targets, we compare the observed ON and OFF light 
profiles of the observed quasars with those of the PSF models 
(Fig.~\ref{fig_results}). The PSF model is 
normalized to match the observed total flux of the quasar. Using 
\textsf{GALFIT} \citep[v.~3.0.2][]{peng02,peng10}, we simulate the light 
profile of a host galaxy of total magnitudes 19, 20, 21 and 22 mag 
(AB system). We assumed a Sersic profile with ellipticity=0.5, $R_e=1''$, 
$0.5''$ and $0.14''$ ($1''\approx5.5$ kpc at the redshift of our targets)
and $n_s$=2. The sampled range of effective radii is defined to 
reproduce the size of the \Lya{} extended emission reported by 
\citet{willott11} in J2329-0301 at $z=6.417$ (diameter of $\sim15$ kpc) and 
more compact CO and [C\,{\sc ii}] emission in the host galaxy of J1148+5251, 
as reported by \citet{walter04} and \citet{riechers09} (CO: 2.5 kpc) and 
\citet{walter09} ([C\,{\sc ii}]: 1.5 kpc). We find that our results are 
practically independent of the ellipticity and the effective radius of the 
host galaxy model for $0.14''<R_e<1''$. 

\subsection{Signatures of gas accretion}\label{sec_accretion}

According to the models by \citet{haiman01}, the \Lya{} emission 
arising from gas surrounding quasar host galaxies at high-$z$ can be as 
bright as $10^{-16}$ erg s$^{-1}$ cm$^{-2}$ arcsec$^{-2}$. 
More recently, \citet{goerdt10} showed that cold gas accreting can give 
rise to significant (up to few times $10^{-17}$ erg s$^{-1}$ cm$^{-2}$ 
arcsec$^{-2}$ for relatively massive galaxies at $z=2.5$) \Lya{} emission 
on 10--100 kpc scales. Such streams are potentially able to survive quasar 
feedback at very high-$z$ \citep{dimatteo12}.

In order to identify any pure line extended emission around the 
quasars, we create `residual' images by subtracting properly scaled OFF 
images from the line+continuum images (`ON'). The scaling is set to match
the total flux observed in the central 5$\times$5 pixels 
($\approx0.2''\times0.2''$, roughly corresponding to the core of the PSF). 
The resulting `residual' images therefore are quasar-subtracted and allow
us to investigate the presence of any extended, pure-line emission around
the quasars (Fig.~\ref{fig_ima}, {\em bottom panels}). 

\subsection{\Lya{} emitting companions}\label{sec_companions}

Finally, we compare the sources detected in the ON images with those in the
OFF images, in order to look for \Lya{} emitters in the field
of our quasars. The FQ889N and FQ906N filters are sensitive to \Lya{} 
emission arising from objects in the redshift ranges $6.279<z<6.355$ and 
$6.415<z<6.492$ respectively, each corresponding to a comoving volume of 
$\approx$0.18 Mpc$^3$ (in the 1 arcmin$^2$ field of view). The 
5-$\sigma$ detection limit for point-sources in these two filters is 
23.43 mag and 23.37 mag ($1.9\times10^{-16}$ erg s$^{-1}$ cm$^{-2}$ and
$1.5\times10^{-16}$ erg s$^{-1}$ cm$^{-2}$) respectively. The observed 
number density of \Lya{} emitters exceeding this flux limit in a blank 
field at $z\sim6.2$ is $\sim10^{-3}$ \Lya{} emitter per arcmin$^2$ 
\citep[e.g.,][]{ouchi10}. This implies that no \Lya{} emitter is expected 
in this volume, for any reasonable overdensity, given our sensitivity.

\begin{figure*}
\begin{center}
\includegraphics[width=0.49\textwidth]{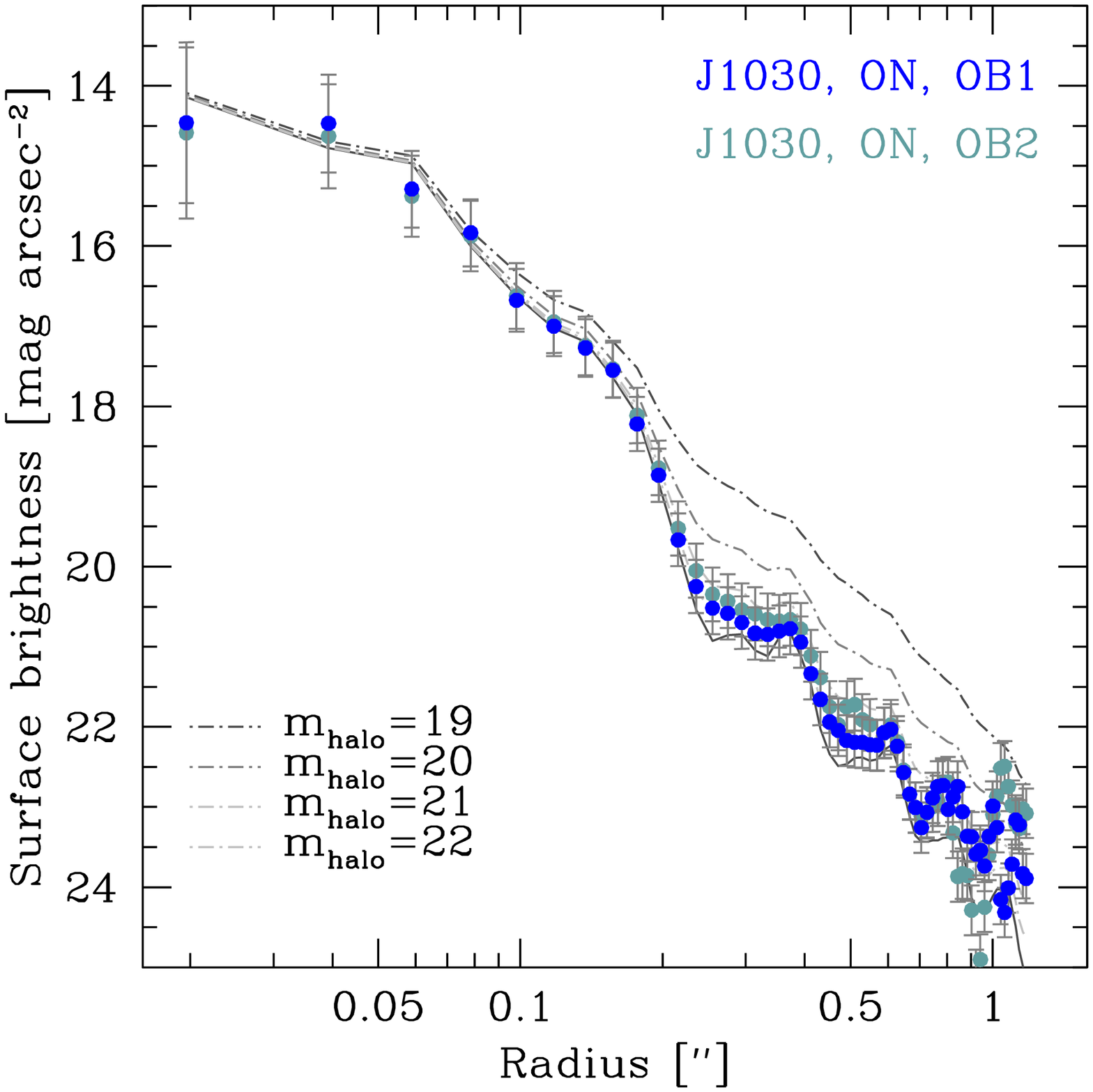}
\includegraphics[width=0.49\textwidth]{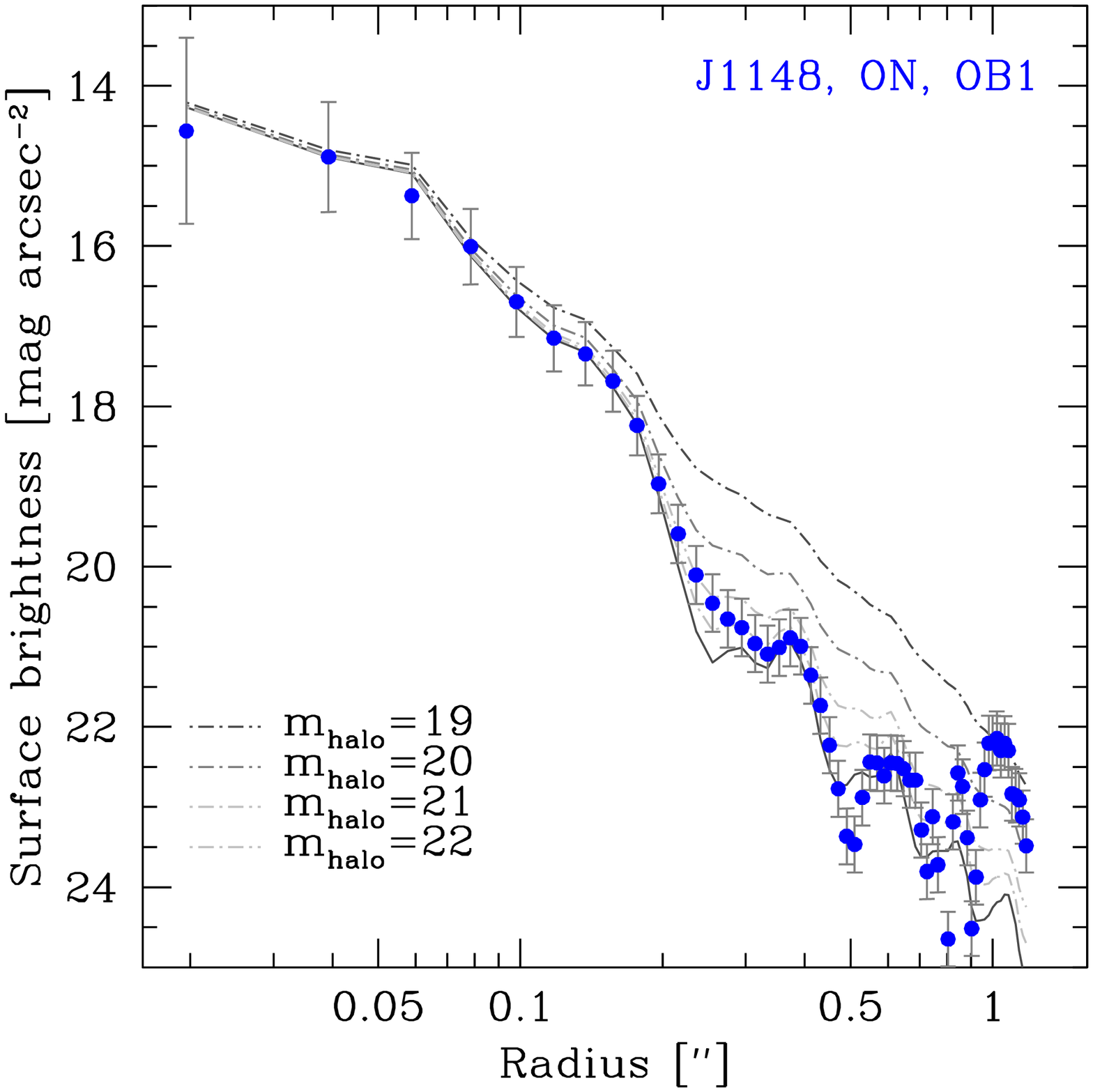}\\
\includegraphics[width=0.49\textwidth]{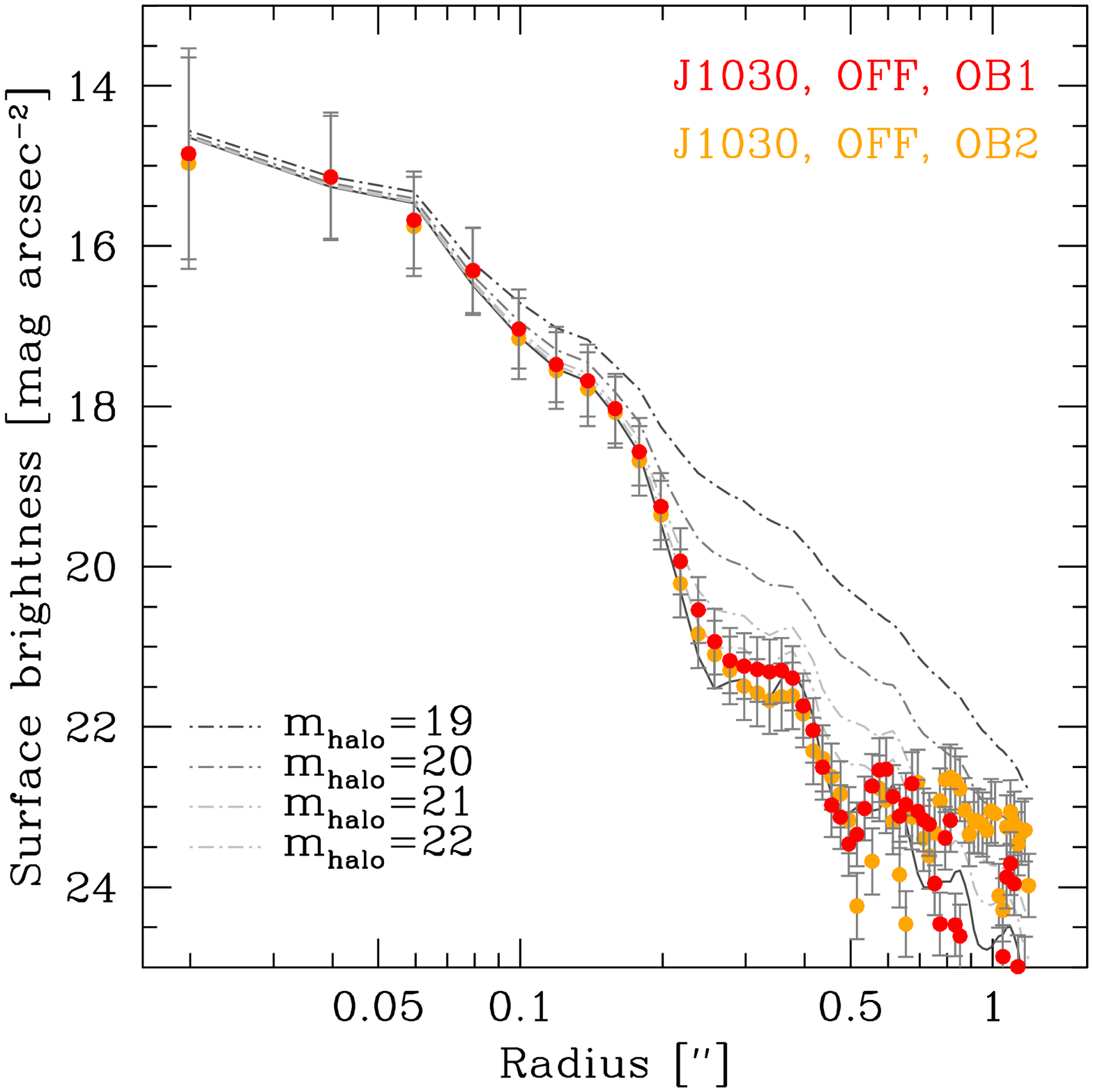}
\includegraphics[width=0.49\textwidth]{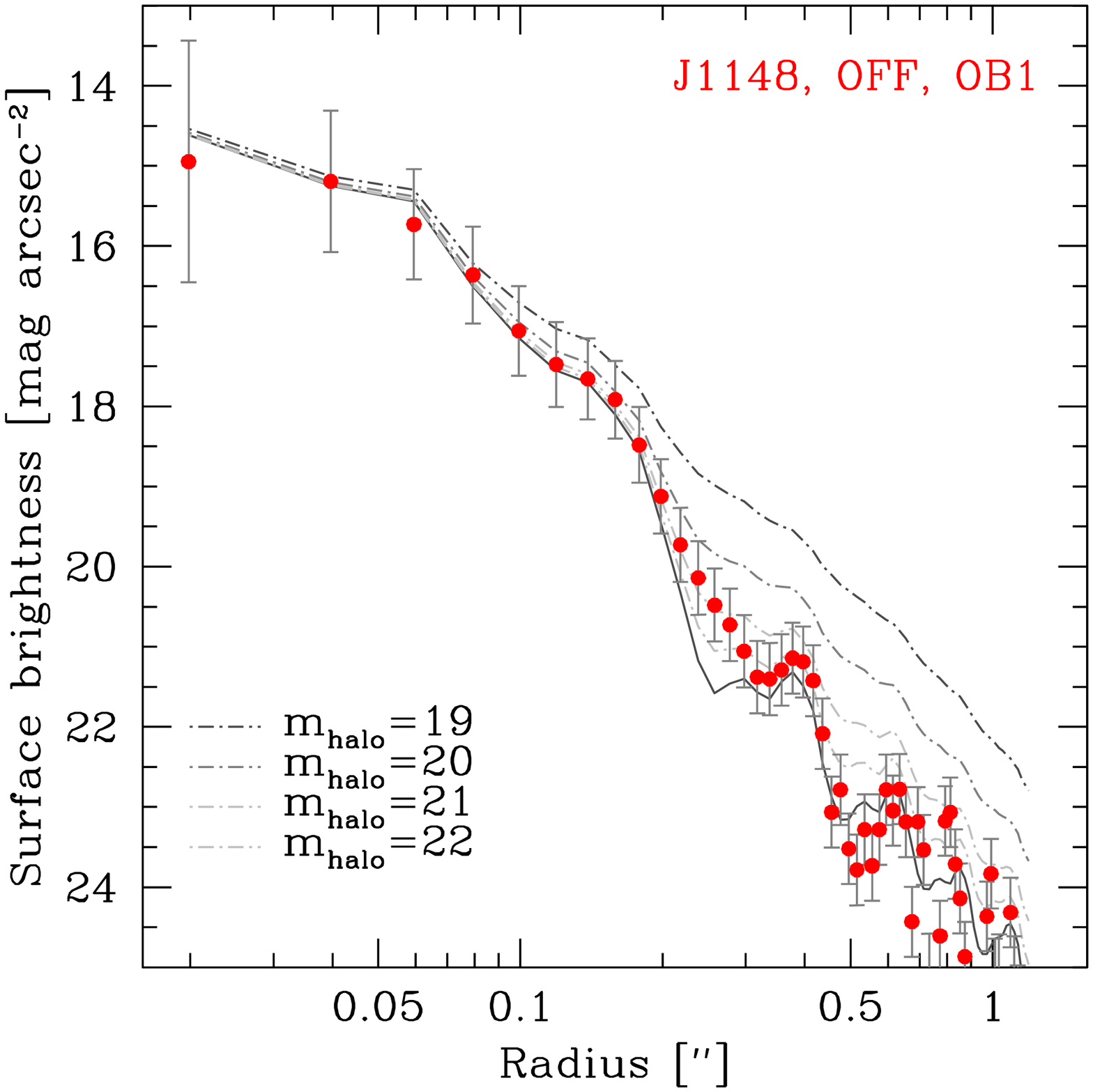}\\
\end{center}
\caption{ON and OFF light profiles of J1030+0524 ({\em left panels})
and J1148+5251 ({\em right panels}). The expected profile for a point-source
plus a host galaxy with a Sersic profile with $R_e=1''$, $n_s=2$, 
ellipticity=0.5, and magnitude (of the extended component only) of 19, 20, 
21 and 22 mag are plotted with dotted lines from dark to light grey. Error
bars are computed as a combination of Poissonian errors, pixelizations and 
background rms. No obvious extended 
emission is observed in any of the panels: the light profiles of the two 
targets are fully consistent with the unresolved, emission from the quasars.
Hosts of total magnitude = 21 would have been detected in our 
observations with $>$2-$\sigma$ significance.}\label{fig_results}
\end{figure*}

\subsection{Results for J1030+0524}\label{sec_j1030}

The light profiles of J1030+0524 (Fig.~\ref{fig_results}) do not show 
any extended component. The 
small light excesses in the OFF observations (of both OBs) at radii 
$\gsim 0.5''$
are due to cosmic ray residuals which are observed in different
positions in the two OBs (see Figure \ref{fig_ima}). We can set an 
upper limit to the magnitude of the extended component by comparing the 
observed profile with the ones simulated using \textsf{GALFIT}. We focus
on the $0.2''-0.8''$ scale, i.e., where the deviations from a PSF are 
expected to be significant but the signal is still high and the PSF models 
are still reliable. Given the uncertainties in the observed light profiles 
and in the PSF model, a host of $\approx21$ mag, both for the line 
and the continuum, is ruled out at 2-$\sigma$. 
Assuming a gaussian model for the \Lya{} line emission, with FWHM=300 \kms{}
and centered at the redshift of the quasar ($z=6.309$, see Figure 
\ref{fig_filters}), we convert the limit set by the ON images into a \Lya{} 
flux limit taking into account the actual throughput curve of the filter and
the redshift of the quasar host. We find $F$(\Lya{},host) $< 7.1 \times
10^{-16}$ erg s$^{-1}$ cm$^{-2}$, i.e., $L$(\Lya{},host) $< 8.3 \times 
10^{10}$ \Lsun{} or $<3.2 \times 10^{44}$ erg\,s$^{-1}$. For the continuum, 
assuming an SED with constant $F_\nu$ for the $k$-correction, we obtain a 
limit on the UV rest-frame luminosity of the host of $M_{1450}>-25.8$ mag. 

A visual inspection of the `Residual' images reveal no significant 
filamentary structure within few arcsec of the source. The comparison 
between the images of the two OBs allow us to discard all the 
low-significance blobs within $5''$ from the quasar as cosmic ray residuals.
We estimate a 5-$\sigma$ surface brightness sensitivity of $\sim1\times 
10^{-17}$ erg s$^{-1}$ cm$^{-2}$ arcsec$^{-2}$ (using a 1 arcsec$^2$ 
aperture). As expected, no \Lya{}
emitter is found in the 1 arcmin$^{2}$ field around the quasar, down to 
a point-source sensitivity of $2.5\times10^{43}$ erg s$^{-1}$ (5-$\sigma$).

\subsection{Results for J1148+5251}\label{sec_j1148}

The light profile of J1148+5251 also does not show any extended component. In
this case a small excess is seen at $\sim1''$ in the ON observation. 
This is most likely due to an observational artifact (see below). 
Following the same approach as adopted for J1030+0524, we can exclude a host 
galaxy brighter than $\approx21$ mag in the ON image (the light profile 
being perfectly consistent with a point-source), and $\approx21$ mag for the
continuum. These limits yield a \Lya{} flux from the host of $< 5.4 \times
10^{-16}$ erg\,s$^{-1}$\,cm$^{-2}$, i.e., $L$(\Lya{},host) $< 6.6 \times 
10^{10}$ \Lsun{} or $<2.5 \times 10^{44}$ erg\,s$^{-1}$. The 
limit on the host continuum is $M_{1450}>-25.8$ mag.

Our observations of J1148+5251 have a depth similar to those of J1030+0524, 
yielding similar limits on the surface brightness ($\sim1\times10^{-17}$ 
erg s$^{-1}$ cm$^{-2}$ arcsec$^{-2}$ at 5-$\sigma$ significance, for a 1 
arcsec$^2$ aperture). A bright 
spot is observed $1''$ South of the quasar in the ON image (see 
Fig.~\ref{fig_ima}), but a careful inspection of the individual frames 
reveals that it is most likely a cosmic ray residual. No other sources 
exceed the 5-$\sigma$ sensitivity limit for point sources (corresponding to
\Lya{} luminosities of $2.6\times10^{43}$ erg s$^{-1}$) within 1 square 
arcmin around the quasar.

\section{Discussion and conclusions}

Our observations set limits on the extended emission of the UV continuum
and of the \Lya{} emission in two quasar host galaxies at $z>6$. The former 
are not very stringent (due to the narrow width of the filters adopted in 
our study). Using the UV continuum luminosity as a probe of star formation 
\citep{kennicutt98}, we obtain a 2-$\sigma$ limit on the UV--based SFR of 
$<1200$ 
\Msun{} yr$^{-1}$, assuming a Salpeter IMF. These UV--based limits are 
in broad agreement with the FIR--based estimates of SFR$\sim$1700--3000 
\Msun{} yr$^{-1}$ reported for J1148+5251 \citep{maiolino05,walter09} if a 
modest extinction correction ($A_{\rm UV}\approx 0.4$ mag) is applied.

On the other hand, the limits on the extended \Lya{} luminosity put tighter 
constraints on the physical properties of our targets. Two obvious sources
of ionizing radiation are present in our targets, namely the accreting black
holes and (at least for J1148+5251) the intense starburst seen at mm 
wavelentghs. Both these processes
are expected to power \Lya{} emission. If \Lya{} emission is powered by the
quasar emission, we can estimate the expected \Lya{} emission from the host
galaxies by modeling the ISM as cold gas clouds absorbing and re-emitting
the light from the quasar. In this scenario, modulo geometrical factors
of the order of unity, and assuming that the ISM clouds are optically thick 
to ionizing photons, the \Lya{} luminosity would be:
\begin{equation}\label{eq_agn_lya}
L({\rm Ly}\alpha)\approx 0.4 f_{\rm c} \, L_{\rm ion.}
\end{equation}
where $f_{\rm c}$ is the covering factor of the clouds and $L_{\rm ion.}$ 
is the ionizing luminosity arising from the black hole accretion 
\citep{hennawi12}. Extrapolating the quasar SED observed in the rest-frame
UV and optical wavelengths using the template by \citet{elvis94}, we 
estimate that $L_{\rm ion.}\approx (3.3-5.4) \times 10^{46}$ erg s$^{-1}$ 
for the two sources. Assuming $f_{\rm c}$=0.1, we infer expected \Lya{} 
luminosities of $\approx (1.3-2.2) \times 10^{45}$ erg s$^{-1}$, i.e., 
one order of magnitude higher than the upper limits set by our observations
(provided that the ISM clouds are distributed over a $\sim$ kpc scale or 
more, i.e., resolved in our observations).

If \Lya{} emission is associated with star formation, we can infer \Lya{}
luminosities from the FIR-based estimates of the SFR  
\citep[through the SFR--\Ha{} relation reported in][]{kennicutt98},
by assuming a standard case B recombination factor of 8.7 for the 
\Lya{}/\Ha{} luminosity ratio,:
\begin{equation}\label{eq_sfr_lya}
\frac{L({\rm Ly}\alpha)}{10^{43} \, \rm erg\ s^{-1}} = 0.11 \ \frac{\rm SFR}{\rm M_\odot\ yr^{-1}}
\end{equation}
In the case of J1148+5251, with a star formation rate of 1700--3000
\citep{maiolino05,walter09}, this implies an expected \Lya{} luminosity of 
$(1.9-3.3)\times10^{45}$ erg s$^{-1}$, i.e., one order of magnitude higher 
than the limit set by our observations.

This difference between expected \Lya{} luminosities and the observational
constraints can be explained by invoking some mechanisms to suppress \Lya{} 
emission. Dust extinction is likely playing a role. A factor $\gsim10$ 
($A_{\rm UV}>2.5$ mag) of extinction is required to explain our limits. Such
a high extinction value is not unexpected in FIR-bright sources,
but is at odds with the relatively low extinction observed towards the
central quasar: \citet{gallerani10} collected low-resolution spectroscopy of
the rest-frame UV emission for a number of high-$z$ quasars, including the
two in our sample, and computed extinction values at 3000 \AA{} (rest 
frame). They find no significant reddening for J1030+0524 and 
$A_{3000}=0.82$ mag (i.e., $A_{\rm UV}\approx1.3$ mag at the wavelengths 
probed in the present study) for J1148+5251.
These relatively modest extinction values, compared 
with the limits set by our observations, suggest a different geometry for 
the highly--opaque dust associated with the kpc-wide starburst and the 
optically--thinner dust along the line of sight to the 
quasar \citep[we note however that FIR-bright quasars tend to have faint 
\Lya{} nuclear emission as well; see, e.g.,][]{wang08b}. Alternatively, 
resonance scattering may prevent \Lya{} emission from
emerging out of the star forming regions. While this effect alone is not
sufficient to explain the lack of strong \Lya{} extended emission, it could 
mitigate the discrepancy if coupled with dust extinction: In this scenario, 
\Lya{} photons from the host repeatedly bounce among optically-thick clouds 
through dusty regions, and get significantly extincted before escaping the
host galaxy. Alternatively, \Lya{} emission may be dim due to a deficit of 
neutral hydrogen around these bright quasars 
\citep[see, e.g.,][]{francis04}. This scenario, however, would be in 
contrast with the large reservoirs of cold gas observed at mm-wavelengths.

It is interesting to compare our limits with the extended \Lya{} emission 
reported around another $z\sim6$ quasar, J2329-0301 ($z=6.417$). 
\citet{goto09} report a diffuse \Lya{} emission of $6.0\times10^{-19}$ 
erg\,s$^{-1}$\,cm$^{-2}$\,\AA$^{-1}$ over an extended region 
($R_e\approx2''$) based on narrow-band imaging with the 8.2m Subaru 
telescope. This implies a diffuse \Lya{} luminosity of 
$3.6\times10^{44}$ erg\,s$^{-1}$, comparable to the limits 
set by our observations\footnote{\citet{goto09} estimate a corresponding 
\Lya{} luminosity of $1.6\times10^{42}$ erg s$^{-1}$, using 
$F$(\Lya)=$F_\lambda$ ($\Delta v/c$) $\lambda_{\rm obs}$ $f$, where 
$\Delta v$=300 km s$^{-1}$ is the line width, $c$ is the speed of light, 
$\lambda_{\rm obs}$ is the observed wavelength of redshifted \Lya{}, and
$f$=60\% is a \Lya{} to total (\Lya{}+cont) correction factor. However,
we point out that in order to retrieve the correct estimate of the \Lya{} 
flux, one should use the filter width ($\approx 1300$ \AA) instead of the
expected line width ($\approx 9$ \AA), making the true flux $\sim100$ times
larger.}. More recently, the same group reported
spectroscopic observations of the same source \citep{goto12}. The extended 
\Lya{} emission has an integrated flux of $(3.6\pm0.2)\cdot10^{-17}$ 
erg\,s$^{-1}$\,cm$^{-2}$, i.e., 20 times fainter than the value reported in 
their imaging observations. \citet{willott11}, using long-slit spectroscopy 
also, found evidence of extended \Lya{} emission around the same source. 
However, they report a {\em lower} limit on the \Lya{} flux of 
$> 1.6 \times 10^{-16}$ erg s$^{-1}$ cm$^{-2}$ (the lower limit is due to 
slit losses and masking of the quasar-dominated area). These last values are 
comparable with the {\em upper} limits set by our observations. Following
the same approach as in \citet{willott11}, we 
re-analysed the Keck HIRES spectra of J1030+0524 and J1148+5251 presented in
\citet{bolton12}. No \Lya{} emission is observed on scales exceeding the 
seeing radius, down to limits comparable with those set by our imaging 
study. If \Lya{} halos were present around the two targets examined in our 
work, they are less prominent than the one reported in J2329-0301.

\section*{Acknowledgments}

We thank the anonymous referee for useful comments which improved the quality
of the manuscript.
We thank C. Leipski and E. Lusso for fruitful discussions on the quasar
SEDs. Support for this work was provided by NASA through grant HST-GO-11640
from the Space Telescope Science Institute, which is operated by AURA, Inc.,
under NASA contract NAS5-26555.RD acknowledges funding from Germany's 
national research centre for aeronautics and space (DLR, project FKZ 50 OR 
1104). XF acknowledge support from NSF grant AST 08-06861 and a David and
Lucile Packard Fellowship. MAS acknowledges support of NSF grant AST-0707266.

\label{lastpage}

\end{document}